\documentclass[12pt]{article}
\begin{document}
\title
{Graphene on noncommutative plane and the Seiberg-Witten map}
\author{
{\bf {\normalsize Aslam Halder}$^{a}
$\thanks{aslamhalder.phy@gmail.com}}\\[0.2cm]
$^{a}${\normalsize Department of Physics, West Bengal State University, Barasat, Kolkata 700126, India}\\
}


\date{}

\maketitle
\begin{abstract}
Graphene on two dimensional (2D) noncommutative (NC) plane in the presence of a constant background magnetic field has been studied. To handel the gauge-invariance issue we start our analysis by a effective massles NC Dirac field theory where we incorporate the Seiberg-Witten (SW) map along with the Moyal star ($\star$) product. The gauge-invariant Hamiltonian of a massless Dirac particle is then computed which is used to study the relativistic Landau problem of graphene on NC plane. Specifically we study the quantum dynamics of a massless relativistic electron moves on monolayer graphene, in the presence of a constant background magnetic field, on NC plane. We also compute the energy spectrum of the NC Landau system in graphene. The results obtained are corrected by the spatial NC parameter $\theta$. Finally we visit the Weyl equation for electron in graphene on NC plane. Interestingly, in this case helicity is found to be $\theta$ modified.    

\end{abstract}

\maketitle

\vskip 1cm
\section{Introduction}
Graphene, 2D configuration of carbon atoms, has drawn lots of interest for research in different branches of physics \cite{novo}-\cite{sharapov}. Remarkably, it is the first actually 2D crystal ever observed in nature. It has a hexagonal honeycomb lattice structure of carbon atoms packed in a 2D system \cite{charles}. 
The Fermi level is situated at the points where the valence band touches the conduction band. These are called the Dirac points $K$ and $K'$. The experimental observations reflects that the charged fermions at the vicinity of the Dirac points show relativistic behavior and they behave like massless Dirac quasi-particle \cite{novo}-\cite{peres}. Theoretically the low energy electronic excitations of graphene can be described by a low energy (2+1) dimensional effective $massless$ Dirac field theory \cite{gw}. Of special attention is paid to the case when this types of planar fermionic system is coupled with external gauge field. In this case the massless Dirac field theory are described by the action 
\begin{eqnarray}
\label{es}
S=\int{d^4x}\bar{\psi}(x)(\gamma^{\mu}\hat{\pi}_{\mu}){\psi}(x)
\end{eqnarray}
where $\psi$ is the Dirac field, $\pi_{\mu}=p_{\mu}-\frac{e}{c}A_{\mu}(x)$ is the gauge-invariant momentum field associated with the field $\psi$ and $A_{\mu}$ is the U(1) gauge field. 
The relativistic behavior of graphene makes it an ideal candidate for the test of quantum field theoretical models. Therefore graphene is a bridge between condensed matter and high energy physics.

On the other hand there has been an upheaval in investigating the physics of systems living in a NC space-time in the last two decads  \cite{sw}-\cite{szabo}. NC space-time arises in string theory with $D$-branes in the background of Neveu-Schwarz fields \cite{acny}-\cite{bcsgscholtz}. It is found that the $D$-brane world volume becomes an NC space and one can arrive at a low energy effective field theory in the point particle limit where the string length goes to zero. This yields a noncommutative quantum field theory (NCQFT) \cite{sw},\cite{scho},\cite{szabo}, \cite{chams}-\cite{Adorno} where the NC coordinate algebra
\begin{eqnarray}
\label{eal}
\left[\hat{X}^{\mu},\hat{X}^{\nu}\right]=i\theta^{\mu\nu}~,
\end{eqnarray}
with the constant anti-symmetric tensor $i\theta^{\mu\nu}$, leads to an uncertainty in the space-time geometry and the notion of a space-time point is replaced by a Planck cell. Besides these, various theories of quantum gravity has also led to NC geometry \cite{suss}-\cite{mof}. 

Surprisingly, in direct analogy with the string theoretical case, noncommuting coordinates can arise in a simple quantum mechanical system. To be specific, 
in the system of charged particle in background EM field, popularly known as the Landau problem \cite{landau}, when the lowest Landau Level is partially filled. Interestingly the conceptual foundation changes drastically in the NC scenario, e.g, the electron is no longer a point like particle and can at best be localized at the scale of the magnetic length. Therefore the study of Landau system in NC space has drawn numerous interest in the literature \cite{jellal}-\cite{ahsg2}.

It has been further noted that a low energy limit of NCQFT leads to noncommutative quantum mechanics (NCQM) which has been studied much in the literature \cite{jellal}
-\cite{sgah}. In this framework the fundamental algebra (which is not the most general form) satisfy by the opetators corresponding to the canonical pairs over 2D NC plane, denoted
by $(\hat{X}_{i},\hat{P}_{i})$ follow:
\begin{eqnarray}
\label{e420}
\left[\hat{X}_{i},\hat{X}_{j}\right]=i\theta_{ij} = i\theta \epsilon_{ij} ~ ; \quad \left[\hat{X}_{i},\hat{P}_{j}\right] = i\hbar\delta_{ij} ~ ;\quad \left[\hat{P}_{i},\hat{P}_{j}\right]=0~.
\end{eqnarray}
Here $\theta$ is the spatial NC parameter and is antisymmetric in the indices $i,j$ as $\theta^{ij}=\theta \epsilon^{ij}$, where $\epsilon^{ij}=-\epsilon^{ji}, (\epsilon^{12}=1)$.
The usual approach in the literature to deal with such problems is to form an equivalent commutative description of the NC theory by employing some transformation which relates the NC operators $\hat{X}_{i}$, $\hat{P}_{i}$ to ordinary commutative operators $\hat{x}_{i}$, $\hat{p}_{i}$ satisfying the usual Heisenberg algebra
\begin{eqnarray}
\left[\hat{x}_{i}, \hat{p}_{j}\right]=i\hbar \delta_{ij}~; \quad \left[\hat{x}_{i} \, , \, \hat{x}_{j}\right]=0= \left[\hat{p}_{i},\hat{p}_{j}\right].
\label{cAlgebra}
\end{eqnarray} 

 All the studies of NCQM give rise to NC corrections to the standard results. Obviously, due to the extreme smallness of the NC parameter $\theta$ it is for sure that its effects will not be observed in the near future. However, there is a nice motivation of studying quantum systems in NC space-time. The physical system in the usual framework can be linked to a noninteracting theory in the NC framework thereby presaging at a possible duality between the two systems \cite{dayijellal}-\cite{gov}. 
 
Besides the study of NCQM in nonrelativistic regime, recently there is a growing interest in the study of relativistic system in NCQM framework in the literature \cite{dmgitman}-\cite{kai}. One of such system is electron in graphene in the presence of a background magnetic field in NC space which is studied by few researchers \cite{dayi}-\cite{vsantos}. However in these studies the issue of NC gauge invariance, which is very crucial for a system with a interacting gauge
field background, sometimes goes unaddressed resulting in a menifestly non-gauge-invariant system. Namely, in \cite{catrina}, \cite{vsantos} where the study of graphene in  NC phase-space is performed, however the authors could not incorporate the effect of spatial
noncommutativity due to the issue of non-gauge-invariance of the Dirac Hamiltonian which would
lead to a gauge-dependent expression for the particle velocity. Therefore it is natural to interrogate whether a gauge-invariant representation of this problem can be possible.    
In the present work we precisely addresses this problem by employing a menifestly gauge-invariant approach. The point of departure of our analysis from the recent papers is a careful attention being paid to the issue of gauge invariance. We incorporate this issue by considering a formalism where the Seiberg-Witten (SW) map \cite{sw} and the Moyal star ($\star$) product \cite{mezin} are used. The SW map is such a map which transforms NC $U(1)_{*}$ gauge system, living in NC space, into the usual $U(1)$ gauge system in the commutative space by preserving the gauge invariance and the physics. This map is obtained by demanding that the ordinary gauge potentials, which are connected by a gauge transformation, are mapped to NC potentials which are connected by the corresponding NC gauge transformation. 

In our formalism the action of the massless NC spinor field coupled with U$(1)_{\star}$ background gauge field in NC space is considered. Applying the SW maps and subsequently expanding the star product and up to first order in the NC parameters $\theta$, we get a menifestly U(1) gauge-invariant commutative equivalent action which describes NC effects as perturbative corrections in terms of commutative fields. By varying this action we acquire the $\theta$ modified massless Dirac equation and subsequintly with a little effort we identify the $\theta$ modified massless Dirac Hamiltonian of the relativistic electron in graphene. This Hamiltonian is manifestly gauge-invariant. With this gauge-invariant Hamiltonian in hand, we then study the relativistic Landau problem of graphene on NC plane. In particular, we compute the time evolution of the position and mechanical momentum operators and found that they pick up the NC correction at the equation of motion level. We then compute the energy spectrum of the system and observed that the spectrum get modified by the spatial NC parameter $\theta$. 
Another important study - the Weyl equation for electron in graphene in this framework, however still lack in the literature, is also considered here. The Weyl equation is obtained from the Dirac equation when the mass of the Dirac particle is zero. Since the electrons in graphene are massless relativistic particle, so it will be interesting to study the Weyl equation for electron in graphene on NC plane for checking whether or not the NC framework alters the eigenvalue of the helicity operator.

The present article is organized as follows. In section 2, we give the basic outline which involves introducing the problem of a massless relativistic charged fermion moving on NC plane in the presence of a constant background EM field first by a NCQM framework and then by a NC field theoretic approach. The SW map in the context of NC gauge-invariance is incorporated here. In section 3, relativistic Landau problem of graphene on NC plane is studied. Specifically, we compute the time evolution of the position and mechanical momentum operators. We also compute the energy spectrum of the relativistic NC Landau system in graphene. We then visit to study the Weyl equation on NC plane in section 4. Finally we summarize our findings in section 5.

\section{Gauge-invariance algorithm in noncommutative space} 
In this section, we want to give a brief overview about the issue of gauge-invariance for the problem of a massless charged fermion coupled with EM field in NC space. First we demonstrate the problem by starting with NCQM description directly where we simply replace the ordinary
product rule among the operators in the quantum mechanical Hamiltonian by Moyal Star ($\star$) product. This yields a matter of non-gauge-invariance. Then we handel this non-gauge-invariance issue by a NC field theoretic approach where we use the SW map along with the Star ($\star$) product and consequently we reach a manifestly gauge-invariant commutative equivalent description of the NC problem.
\subsection{Issue of gauge-non-invariance}
We consider a relativistic massless charged fermion of charge $e$ moving on NC plane in the presence of a background EM field. 
The quantum dynamics of this particle can be desceibed by the generalization of the standard Dirac equation for such particle in
the commutative space to the NC space as 
\begin{eqnarray}
\label{evx}
i\hbar\frac{\partial\Psi(X)}{\partial t}=\hat{H}(X, P)\star\Psi(X)
\end{eqnarray}
where
\begin{eqnarray}
\label{evxy}
\hat{H}(X,P)=c\vec{\alpha}\cdot\left(\vec{\hat{P}}-\frac{e}{c}\vec{\hat{\mathcal{A}}}(X)\right)+e\phi(X)
\end{eqnarray}
is the NC Dirac Hamiltonian of the system, ${\mathcal{A}}(X)$ and $\phi(X)$ are the EM vector potential and scalar potential respectively in NC space, $\Psi(X)$ is the 4-component NC spinor of the system. It should be noted that we make the NC generalization by replacing the ordinary product between the functions of commutative variables with the Moyal star ($\star$) product between the functions of NC variables. The Moyal star product is defined as
\begin{eqnarray}
\label{ev03}
f(X)\star g(X)=f(x)exp\left\{\frac{i}{2}\theta^{\mu\nu}\overleftarrow{\partial_{\mu}}\overrightarrow{\partial_{\nu}}\right\}g(x)
\end{eqnarray}
with $f(X)$, $g(X)$ are arbitrary functions of NC variables whereas $f(x)$, $g(x)$ are functions of commutative variables. Appling the star product in eq. (\ref{evx}) we get  the NC corrected (up to first order in $\theta$) Dirac equation as
\begin{eqnarray}
\label{ej467}
i\hbar\frac{\partial\Psi(x)}{\partial t}&=&\hat{H}(x, p)\Psi(x)+\frac{i}{2}\theta^{jk}\partial_{j}\hat{H}(x, p)\partial_{k}\Psi(x)\nonumber\\
&=&\hat{H}(x, p)\Psi(x)+\frac{i}{2}\theta^{jk}\partial_{j}\left\{c\alpha_{j}\left(\hat{p}_{j}-\frac{e}{c}\hat{\mathcal{A}}_{j}(x)\right)+e\phi(x)\right\}\partial_{k}\Psi(x)~.
\end{eqnarray}
From the above equation we can easily identify the $\theta$ modified Dirac Hamiltonian of the massless relativistic charged fermion moving on NC plane in the presence of constant background EM field\footnote{$\theta^{i}=\frac{1}{2}\epsilon_{ijk}\theta^{jk}$}
\begin{eqnarray}
\label{es04}
\hat{H}=c\vec{\alpha}\cdot\left(\vec{\hat{p}}-\frac{e}{c}\vec{\hat{\mathcal{A}}}(x)\right)+e\phi(x)+\frac{e}{2\hbar}\left\{\vec{\bigtriangledown}\left(c\vec{\alpha}\cdot\vec{\hat{\mathcal{A}}}(x)-\phi(x)\right)\times\vec{\hat{p}}\right\}\cdot\vec{\theta}~.
\end{eqnarray}

The above Hamiltonian is not manifestly gauge invariant due to the presence of gauge dependant term in the $\theta$ correction portion. Therefore we can not use this Hamiltonian to study the relativistic Landau problem of graphene. 
 In \cite{bertolami}, \cite{catrina} this gauge symmetry breaking issue is appeared. To avoid this crux the authors renounce the spatial noncommutativity and consider only momentum noncommutativity in their work. This matter shed light on the fact that when we are working with a system of charged particle in NC space and the system is coupled with a NC gauge field, the NC quantum mechanical treatment (considering only star product) of the problem is an incomplete description in conformity with NC gauge invariance.

\subsection{Gauge-invariant nocommutative massless Dirac field theory} 
To tackle the issue of gauge-non-invariance we want to use a field theoretic approach in the problem where we will generalize the system from commutative space to
NC space and finally sticking to a commutative equivalent description of the NC system.
The NC generalization is done through replacing the ordinary commutative spinor field $\psi$ and U(1) gauge field $A_{\mu}$ in the action (\ref{es}) by the  NC spinor field ${\Psi}(X)$ and U$(1)_{\star}$ gauge field ${\mathcal{A}}^{\mu}(X)=({\mathcal{A}}^{0}, {\mathcal{A}}^{i});~i=1,2,3$. Also the ordinary product should be replaced by the Moyal star product. The corresponding U$(1)_{\star}$ gauge invariant action is  
\begin{eqnarray}
\label{eur}
S&=&\int{d^4x}\bar{\Psi}(X)\star(\gamma^{\mu}\hat{\Pi}_{\mu})\star{\Psi}(X)
\end{eqnarray}
where $\Pi_{\mu}=P_{\mu}-\frac{e}{c}\mathcal{A}_{\mu}(X)$ is the NC generalization of the momentum field $\pi_{\mu}$. 
The NC gauge field ${\mathcal{A}}_{\mu}$ and the NC spinor field ${\Psi}$ are expressed in terms of ordinary commutative gauge field $A_{\mu}$ and spinor field $\psi$ by the SW map (up to first order in $\theta$) \cite{sw}
\begin{eqnarray}
\label{e159}
{\mathcal{A}}_{\mu}&=&{A}_{\mu}+\frac{e}{2\hbar c}\theta^{\alpha\beta}{A}_{\alpha}(\partial_{\beta}{A}_{\mu}+F_{\beta\mu})\\
{\Psi}&=&\psi+\frac{e}{2\hbar c}\theta^{\alpha\beta}{A}_{\alpha}\partial_{\beta}\psi~.\nonumber
\end{eqnarray}
Applying the above map and subsequently using the Moyal star product in the action (\ref{eur}), we compute the $\theta$ modified commutative equivalent action as 
\begin{eqnarray}
\label{e163vx}
S^{\theta}&=&\int d^4x\bar{\psi}(x)\left\{\gamma^{\mu}\left[\left(1+\frac{e}{4c\hbar}\theta^{\alpha\beta}F_{\alpha\beta}\right)\hat{\pi}_{\mu}- \frac{e}{2c\hbar}\theta^{\alpha\beta}F_{\alpha\mu}\hat{\pi}_{\beta}\right]\right\}\psi(x)~.
\end{eqnarray}
The above action is manifestly U(1) gauge invariant due to the appearence of the of commutative gauge
field strength tensor $F_{\alpha\beta}$ and commutative momentum field $\pi_{\mu}$. It should be noted that incorporating of the SW map along with the star product in our work confirm the gauge-invariance of the above action.  

We now vary the above action to achive the $\theta$ modified Dirac equation which reads\footnote{We only consider spatial noncommutativity and ignore time-space noncommutativity, $\theta^{0\mu}=0$} 
\begin{eqnarray}
\label{e163x9}
i\hbar\partial_{t}\psi=\hat{\uppercase{H}}\psi
\end{eqnarray}
where the Hamiltonian $\hat{\uppercase{H}}$ in the above equation is given by
\begin{eqnarray}
\label{eHm}
\hat{\uppercase{H}}&=&(\hat{H}_{D}+\Delta\hat{H}^{\theta})\\
\hat{H}_{D}&=&c\vec{\alpha}\cdot\left(\vec{\hat{p}}-\frac{e}{c}\vec{\hat{A}}\right)+eA_{0}\nonumber\\
 \Delta\hat{H}^{\theta}&=&\frac{e}{2\hbar }\left\{([\overrightarrow{E}\times\overrightarrow{\hat{\Pi}}]\cdot\vec{\theta})+[\overrightarrow{\theta}\times(\overrightarrow{\alpha}\times\overrightarrow{B})]\cdot\vec{\hat{\Pi}}\right\}\nonumber~.
\end{eqnarray}
As because the above Hamiltonian $\hat{\uppercase{H}}$ is derived from the manifestly gauge-invariant action (\ref{e163vx}), so we can claim without any ambiguity that $\hat{\uppercase{H}}$ is also gauge-invariant. We clearly observe that there is no gauge dependent term present in the above Hamiltonian which is totally contrast with the non-gauge-invariant Hamiltonian (\ref{es04}). Therefore we can use this gauge-invariant Hamiltonian to study the relativistic Landau problem of graphene in the next section.


\section{Relativistic Landau problem of graphene in noncommutative plane }
We are now in a position to study the NC relativistic landau problem of graphene, more elaborately the quantum dynamics of a massless relativistic electron moves on a monolayer graphene sheet on NC plane in the presence of a background homogeneous magnetic field perpendicular to the plane.  
To do this task we choose specific direction for the magnetic field as $\vec{B} = B\hat{k}$ so that our system is confined to the NC $x$-$y$ plane. 
With this setting and by a little manipulation of the gauge-invariant Hamiltonian (\ref{eHm}), we can write the Hamiltonian of the Landau system of graphene as \footnote{Since $\theta^{\mu\nu}=0$ in the present work, therefore we don't consider the electric field in the subsequent part of the paper.}
\begin{eqnarray}
\label{e163kj}
\hat{H}&=&v_{F}\vec{\alpha}\cdot\left(\vec{\hat{p}}-\frac{e}{c}\vec{\hat{A}}\right)\left(1+\frac{eB\theta}{2\hbar c}\right)\nonumber\\
&=&v_{F}\left\{\alpha_{x}\left(\hat{p}_{x}+\frac{eB\hat{y}}{2c}\right)+\alpha_{y}\left(\hat{p}_{y}-\frac{eB\hat{x}}{2c}\right)\right\}\left(1+\frac{eB\theta}{2\hbar c}\right)
\end{eqnarray}
where we have fixed the symmetric gauge choice for the vector potential $\vec{A}\equiv\left(-\frac{B\hat{x}}{2},\frac{B\hat{x}}{2},0\right)$ and replace the speed of light in vacuum ($c$) by the Fermi velocity ($v_{F}=10^{6}$) with which the relativistic massless electron in graphene moves. 
It should be noted that the manifest gauge-invariance of the commutative equivalent action (\ref{e163vx}) confirm that at all the
subsequent levels (i.e., the equation of motion or the Hamiltonian etc) our theory remains gauge-invariant
and therefore we can choose this gauge without loosing generality.

Let us study the relativistic dynamics of the electron in graphene in the presence of a background magnetic field on NC plane. The time evaluation of the position operator (we deliberately refer this quantity as the velocity operator in the reminder) is computed by using the Hamiltonian (\ref{e163kj}) as 
\begin{eqnarray}
\label{e4x} 
\vec{v}^{NC}&=&\frac{i}{\hbar}[\hat{H}~, ~\hat{i}\hat{x}+\hat{j}\hat{y}]\nonumber\\
&=&\frac{i}{\hbar}\left[v_{F}\left\{\alpha_{x}\left(\hat{p}_{x}+\frac{eB\hat{y}}{2c}\right)+\alpha_{y}\left(\hat{p}_{y}-\frac{eB\hat{x}}{2c}\right)\right\}\left(1+\frac{eB\theta}{2\hbar c}\right)
,~\hat{i}\hat{x}+\hat{j}\hat{y}\right]\nonumber\\
&=&v_{F}\vec{\alpha}\left(1+\frac{eB\theta}{2\hbar c}\right)~.
\end{eqnarray}
The time-evolution of the gauge-invariant
mechanical momentum operator $\vec{\hat{\pi}}$ (which is analogue to the classical Lorentz force) in NC space is obtained as
\begin{eqnarray}
\label{e4xgy} 
\frac{d}{dt}(\vec{\hat{\pi}})\equiv\vec{F}^{NC}&=&\frac{i}{\hbar}[\hat{H}~,~\vec{\hat{\pi}}]\nonumber\\
&=&\frac{i}{\hbar}[\hat{H}~,~\hat{i}\hat{\pi}_{x}+\hat{j}\hat{\pi}_{y}]\nonumber\\
&=&\frac{i}{\hbar}\left[v_{F}\left\{\alpha_{x}\left(\hat{p}_{x}+\frac{eB\hat{y}}{2c}\right)+\alpha_{y}\left(\hat{p}_{y}-\frac{eB\hat{x}}{2c}\right)\right\}\left(1+\frac{eB\theta}{2\hbar c}\right)\right.\nonumber\\
&&\left.~,~\hat{i}\left(\hat{p}_{x}+\frac{eB\hat{y}}{2c}\right)+\hat{j}\left(\hat{p}_{y}-\frac{eB\hat{x}}{2c}\right)\right]\nonumber\\
&=&\frac{1}{c}e\vec{v}^{NC}\times\vec{B}~.
\end{eqnarray}
From the expressions (\ref{e4x}) and (\ref{e4xgy}) we observe that the velocity of the electron in graphene on NC plane and the Lorentz force on it 
pick up the NC correction. Notice that the NC effect arise completely at the equation of
motion level. Since equation of motion determines the experimentally observable dynamics of any
physical system so we can safely assume that any observable NC quantum machanical effect displayed by our
system can be captured by replacing the standard velocity and Lorentz force by the effective velocity and Lorentz force. This is a crucial
result of our paper. At $\theta=0$ limit, the above expressions for the velocity and the
Lorentz force are exactly match with their commutative forms.

We now turn to compute the energy spectrum of the Landau system in graphene on NC plane. To do so we recast the Hamiltonian (\ref{e163kj}) into following form
\begin{eqnarray}
\label{e10s}
\hat{H}&=&v_{F}\vec{\alpha}\cdot\vec{\hat{\pi}}\left(1+\frac{eB\theta}{2\hbar c}\right)\nonumber\\
&=&v_{F}\vec{\alpha}\cdot\tilde{\vec{\hat{\pi}}}\nonumber\\
&=&v_{F}\left(
\begin{array}{cc}
\vec{\sigma}\cdot\tilde{\vec{\hat{\pi}}}&0  \\
0&-\vec{\sigma}\cdot\tilde{\vec{\hat{\pi}}} \\
\end{array}
\right)\nonumber\\
&=&v_{F}\left(
\begin{array}{cc}
\hat{H}^k&o  \\
0&\hat{H}^{k^{'}} \\
\end{array}
\right)
\end{eqnarray}
where we define $\tilde{\hat{\pi}}=\hat{\pi}\left(1+\frac{eB\theta}{2\hbar c}\right)$ as the NC modified mechanical momentum operator. Further we have applied the usual representation of the Dirac matrix $\alpha$. The Hamiltonian (\ref{e10s}) represent two copies of Hamiltonian for massless relativistic elecrtons around each Dirac points $k$ and $k^{'}$ in the two dimensional graphene crystal on NC plane in the presence of a homogeneous background magnetic field.
For the Dirac point $K$, the single copy of Hamiltonian (\ref{e10s}) is given by
\begin{eqnarray}
\label{e163k5w}
\hat{H}^k=v_{F}\vec{\sigma}\cdot\tilde{\vec{\hat{\pi}}}~.
\end{eqnarray}
We want to evaluate the energy eigenvalue of the above Hamiltonian. In order to do this let us first compute the commutator bracket between the two components of NC modified mechanical momenum $\tilde{\vec{\pi}}$ operator, which yields 
\begin{eqnarray}
\label{eo0w32}
[\tilde{\pi}_{x}~,~\tilde{\pi}_{y}]&=&\left(1+\frac{eB\theta}{2\hbar c}\right)^2\left[\hat{p}_{x}-\frac{e}{c}\hat{A}_{x}~,~\hat{p}_{y}-\frac{e}{c}\hat{A}_{y}\right]\nonumber\\
&=&\left(1+\frac{eB\theta}{2\hbar c}\right)^2\left[\hat{p}_{x}+\frac{eB\hat{y}}{2c}~,~\hat{p}_{y}-\frac{eB\hat{x}}{2c}\right]\nonumber\\
&=&i\frac{\hbar^2}{l_{B}^2}\left(1+\frac{eB\theta}{2\hbar c}\right)^2\nonumber\\
&=&i\frac{\hbar^2}{\tilde{l}_{B}^2}
\end{eqnarray}
where $l_{B}=\sqrt{\frac{\hbar c}{eB}}$ is the magnetic length which is the characteristic length scale for the problem
of a charged particle moves in a constant background magnetic field and $\tilde{l}_{B}=l_{B}\left(1+\frac{eB\theta}{2\hbar c}\right)^{-1}$ is the NC corrected magnetic length. 

Now we define the following ladder operators in terms of the NC modified magnetic length $\tilde{l}_{B}$ and the NC modified mechanical momentum operator $\tilde{\hat{\pi}}$
\begin{eqnarray}
\label{eo0bg2}
\hat{a}&=&\frac{\tilde{l}_{B}}{\sqrt{2}\hbar}(\tilde{\hat{\pi}}_{x}+i\tilde{\hat{\pi}}_{y})\nonumber\\
\hat{a}^{\dagger}&=&\frac{\tilde{l}_{B}}{\sqrt{2}\hbar}(\tilde{\hat{\pi}}_{x}-i\tilde{\hat{\pi}}_{y})~.
\end{eqnarray}
With the help of eq. (\ref{eo0w32}), we can show that the ladder operators $\hat{a}$ and $\hat{a}^{\dagger}$ satisfy the following commutation relation
\begin{eqnarray}
\label{ebr}
[\hat{a}~,~\hat{a}^{\dagger}]=1~.
\end{eqnarray}
In terms of the ladder operators $\hat{a}$ and $\hat{a}^{\dagger}$, the Hamiltonian (\ref{e163k5w}) is expressed as
\begin{eqnarray}
\label{ebr1}
\hat{H}^k=\tilde{\omega}\hbar\left(
\begin{array}{cc}
0&\hat{a}^{\dagger}  \\
\hat{a}&0 \\
\end{array}
\right)
\end{eqnarray}
where $\tilde{\omega}=\frac{\sqrt{2}v_{F}}{\tilde{l}_{B}}$ is the NC modified relativistic cyclotron frequency.

The eigenvalue equation for the Hamiltonian $\hat{H}^k$ is given by
\begin{eqnarray}
\label{ebrc}
\hat{H}^{k}\psi^k&=&E^{k}\psi^k\nonumber\\
\tilde{\omega}\hbar\left(
\begin{array}{cc}
0&\hat{a}^{\dagger}  \\
\hat{a}&0 \\
\end{array}
\right)\left(
\begin{array}{cc}
\psi^{k}_{A}  \\
\psi^{k}_{B} \\
\end{array}
\right)&=&E^{k}\left(
\begin{array}{cc}
\psi^{k}_{A}  \\
\psi^{k}_{B} \\
\end{array}
\right)
\end{eqnarray}
where $\psi^k$ is the the 2-component spinor for the electronic state at the Dirac point $K$, $E^k$ is the energy eigenvalue of the Hamiltonian $\hat{H}^{k}$, $A$ and $B$ are the two sub-lattices for each Dirac point.
From the above eigenvalue equation we get two coupled equations as
\begin{eqnarray}
\tilde{\omega}\hbar \hat{a}^{\dagger}\psi^{k}_{B}=E\psi^{k}_{A}\\
\tilde{\omega}\hbar \hat{a}\psi^{k}_{A}=E\psi^{k}_{B}~.
\end{eqnarray}
We eleminate $\psi^{k}_{B}$ from the above coupled equations and get the following equation for $\psi^{k}_{A}$ as
\begin{eqnarray}
\label{e54kh9x}
\hat{a}^{\dagger}\hat{a}\psi^{k}_{A}=\left(\frac{E}{\tilde{\omega}\hbar}\right)^{2}\psi^{k}_{A}~.
\end{eqnarray}
Since the ladder operators follow the commutation relation (\ref{ebr}), so we can treat $\hat{a}^{\dagger}\hat{a}$ as a number operator which obeys the folllowing eigenvalue equation
\begin{eqnarray}
\label{edt4x}
\hat{a}^{\dagger}\hat{a}|n>&=&n|n>~,~n=0,1,2,3,\cdots
\end{eqnarray}
with
\begin{eqnarray}
\label{e54kh9xr}
\hat{a}^{\dagger}|n>&=&\sqrt{n+1}|n+1>\nonumber\\
\hat{a}|n>&=&\sqrt{n}|n-1>\nonumber~.
\end{eqnarray}
Finally comparing eq. (\ref{e54kh9x}) with eq. (\ref{edt4x}) we obtain the energy spectrum of the relativistic Landau problem of graphene on NC plane as
\begin{eqnarray}
\label{ev3r}
E&=&\pm\tilde{\omega}\hbar\sqrt{n}\nonumber\\
&=&\pm\frac{\hbar v_{F}}{l_{B}}\sqrt{2n}\left(1+\frac{eB\theta}{2c\hbar}\right)~.
\end{eqnarray} 
The above result clearly reveals that the relativistic landau levels are altered by spatial noncommutativity. At $\theta=0$ limit, the result returns to its commutative form. It should be noted that even in the first order approximation we get a NC correction coming from the spatial sector of the NC algebra which is in contrast with the result obtained in \cite{catrina} where the NC correction of the Landau levels of graphene is independent of spatial NC parameter $\theta$.

\section{Weyl equation in noncommutative plane and the modified helicity eigenvalue}
Having further investigation of the aspects of the spatial noncommutativity, in this section we want to achieve some insight into its possible impression in another problem related to the massless Dirac equation, namely the Weyl equation. Before going to study the NC picture of the
problem we first want to present a brief review of its ordinary commutative version. In commutative space the Dirac equation for massive relativistic particle reads
\begin{eqnarray}
\label{evh}
i\hbar\partial_{t}\psi=\left(c\vec{\alpha}\cdot\vec{\hat{p}}+\beta mc^2\right)\psi~.
\end{eqnarray}
Now if we express the 4-component spinor $\psi$ in terms of 2-component spinors $\chi_{+}$ and $\chi_{-}$ as $\psi=\left(
\begin{array}{cc}
\chi_{+} \\
\chi_{-} \\
\end{array}
\right)$, then the 2-component spinors $\chi_{\pm}$ are coupled with each other due to the presence of mass term in the Dirac equation and we have the coupled Dirac equations which read 
\begin{eqnarray}
\label{evg}
i\hbar\partial_{t}\chi_{+}&=&c\vec{\sigma}\cdot\vec{\hat{p}}\chi_{+}+mc^2\chi_{-}\nonumber\\
i\hbar\partial_{t}\chi_{-}&=&-c\vec{\sigma}\cdot\vec{\hat{p}}\chi_{-}+mc^2\chi_{+}~.
\end{eqnarray}
In the limit when the mass of the Dirac particle goes to zero, the coupling between the 2-compont spinors $\chi_{\pm}$ is break and we get the following uncoupled equations
\begin{eqnarray}
\label{evg1}
i\hbar\partial_{t}\chi_{+}&=&c\vec{\sigma}\cdot\vec{\hat{p}}\chi_{+}\nonumber\\
i\hbar\partial_{t}\chi_{-}&=&-c\vec{\sigma}\cdot\vec{\hat{p}}\chi_{-}
\end{eqnarray} 
which are popularly known as Weyl equations and the 2-component spinors $\chi_{\pm}$ are called the Weyl spinors. We recast the above Weyl equations into following compact form
\begin{eqnarray}
\label{evg1}
\hat{p}_{0}\chi_{\pm}&=&\pm\vec{\sigma}\cdot\vec{\hat{p}}\chi_{\pm}\nonumber\\
\frac{\vec{\hat{p}}\cdot\vec{\sigma}}{2|p_{0}|}\chi_{\pm}&=&\pm\frac{1}{2}\chi_{\pm}
\end{eqnarray}
where $\hat{p}_{0}\equiv i\hbar\frac{1}{c}\partial_{t}$. 
Since $\frac{\vec{\hat{p}}\cdot\vec{\sigma}}{2|p_{0}|}$ is the helicity operator, therfore the above equations give us a very important message that the Weyl spinors $\chi_{\pm}$  are eigensates of the helicity operator $\frac{\vec{\hat{p}}\cdot\vec{\sigma}}{2|p_{0}|}$ with eigenvalue $\pm\frac{1}{2}$.
 
If there is an external EM interaction in the system, then we should make the substitution $p_{\mu}\rightarrow\pi_{\mu}\equiv(\pi_{0},\vec{\pi})=p_{\mu}-\frac{e}{c}A_{\mu}$. 
Therefore in this case eq. (\ref{evg1}) becomes
\begin{eqnarray}
\label{evg2}
i\hat{\pi}_{0}\chi_{\pm}&=&\pm\vec{\sigma}\cdot\vec{\hat{\pi}}\chi_{\pm}~.
\end{eqnarray}   
The above equation clearly implies that the interaction of charged spin-$\frac{1}{2}$ relativistic particle, for example an relativistic electron, with the external EM field does not introduce any coupling between the two Weyl spinors $\chi_{\pm}$, the only coupling is because of particle's mass. Physically this signifies that the helicity is a constant of motion for the system of an ultrarelativistic Dirac particle moving in an external EM background when the energy of the system is very high compared to its rest mass.

 Concerning to the above point, the relativistic electron present in graphene is a massless particle. So the coupling between the Weyl spinors is naturally absent here whether the system is in external EM background or not. However the inherent query is that when graphene, interacting with external EM field, is in NC space what will happen. Surprisingly such study is still lacking in the literature. We are going to investigate this striking matter in this section. 

We start with the commutative equivalent description of NC Dirac equation for a massless relativistic electron in graphene, interacting with a constant background magnetic field, on two dimensional NC plane which reads
\begin{eqnarray}
\label{evh1}
i\hbar\partial_{t}\psi&=&\hat{H}\psi\nonumber\\
i\hbar\partial_{t}\left(
\begin{array}{cc}
\chi_{+} \\
\chi_{-} \\
\end{array}
\right)&=&\left(v_{F}\vec{\alpha}\cdot\tilde{\vec{\hat{\pi}}}\right)\left(
\begin{array}{cc}
\chi_{+} \\
\chi_{-} \\
\end{array}
\right)
\end{eqnarray}
where we have used the Hamiltonian (\ref{e10s}). Due to the absence of mass term in the above equation it is clearly seen that there is no coupling between the Weyl spinors $\chi_{\pm}$. Therefore from the above Dirac equation basically we get two uncoupled Weyl equations in NC plane as 
\begin{eqnarray}
\label{e1u}
i\hbar\partial_{t}\chi_{+}&=&v_{F}\vec{\sigma}\cdot\tilde{\vec{\hat{\pi}}}\chi_{+}\nonumber\\
i\hbar\partial_{t}\chi_{-}&=&-v_{F}\vec{\sigma}\cdot\tilde{\vec{\hat{\pi}}}\chi_{-}~.
\end{eqnarray} 
Now we recast the above two NC Weyl equations into the following compact form
\begin{eqnarray}
\label{e10u}
i\hbar\partial_{t}\chi_{\pm}&=&\pm v_{F}\vec{\sigma}\cdot\tilde{\vec{\hat{\pi}}}\chi_{\pm}\nonumber\\
\hat{\pi}_{0}\chi_{\pm}&=&\pm\vec{\sigma}\cdot\vec{\hat{\pi}}\left(1+\frac{eB\theta}{2\hbar c}\right)\chi_{\pm}\nonumber\\
\frac{\vec{\sigma}\cdot{\vec{\hat{\pi}}}}{2|\hat{\pi}_{0}|}\chi_{\pm}&=&\pm\frac{1}{2}\left(1+\frac{eB\theta}{2\hbar c}\right)^{-1}\chi_{\pm}\nonumber\\
&=&\pm\frac{1}{2}\left(1-\frac{eB\theta}{2\hbar c}+\mathcal{O}(\theta^2)+\cdots\right)\chi_{\pm}~.
\end{eqnarray}
From the above equation we observe that when graphene is on NC plane in the presence of a constant background magnetic field, the eigenvalue of the helicity operator is modified by the spatial NC parameter $\theta$. This observation clearly reavels that though the helicity is a constant of motion for a massless Dirac particle interacting with external EM field in commutative space, however in NC space it is no longer a constant of motion. This is an important finding by us. At $\theta=0$ limit, we recover the commutative eigenvalue of the helicity operator.

\section{Conclusions}
In this paper we have studied the relativistic quantum dynamics of an electron in monolayer graphene, interacting with a constant background magnetic field, on NC plane. First we have started with NCQM description of the problem directly where one simply replaces the ordinary product rule among the quantum mechanical operators by the Moyal star ($\star$) product and thereby a manifestly non-gauge-invariant commutative equivalent Hamiltonian of the NC system is obtained. To handel the gauge-invariance issue we then consider a NC field theoretic approach. In this approach we restart with the action
of a NC spinor field coupled with U$(1)_{\star}$ gauge field.
Expanding the star product and subsequently employing the SW maps up to first order in the NC parameter
$\theta$, we arrive at a menifestly U(1) gauge-invariant action which describes NC effects as
perturbative corrections in terms of commutative fields. Varying the commutative equivalent action we obtain the NC Dirac equation and subsequently identify the NC Dirac Hamiltonian. With a little manipulation we then obtain the NC Dirac Hamiltonian of a relativistic massless electron, interacting with a constant background magnetic field, in graphene on NC plane. With the Hamiltonian in hand, we then go to study the quantum dynamics of the electron in graphene. In particular, we compute the time evolution of the position and mechanical momentum operators. The results obtained are found to get NC corrected. The main result in the paper is the computation of the energy spectrum of the NC Landau system of graphene. We observe that the Landau levels are altered by $\theta$. Finally we investigate the Weyl equation for electron in graphene on NC plane. In this case we evaluate the eigenvalue of the helicity operator and found that the eigenvalue pick up the NC correction. This tells that the helicity does not remain a constant of motion in NC space.

\section*{Acknowledgement}
The author would like to thank Anirban Saha for important discussions on the matter of this work.

\end{document}